\documentclass[prb,a4paper,12pt,onecolumn,superscriptaddress,amsmath,amsfonts,amssymb,preprintnumbers,citeautoscript]{revtex4}

\pdfoutput=1
\usepackage{graphicx,color,booktabs,microtype,afterpage,sidecap}
\usepackage[charter]{mathdesign}

\usepackage[colorlinks,plainpages=false,linkcolor=black,urlcolor=blue,citecolor=black,pdfpagemode=UseNone,pdfstartview=FitBH]{hyperref}

\makeatletter
\renewcommand{\@evenfoot}{\hfill\raisebox{-3em}{\bf\thepage}\hfill}
\renewcommand{\@oddfoot}{\hfill\raisebox{-3em}{\bf\thepage}\hfill}
\makeatother

\begin{document}

\title{Long-range transfer of electron-phonon coupling in oxide superlattices}
\author{N.~Driza}
\affiliation{Max-Planck-Institut~f\"{u}r~Festk\"{o}rperforschung,
Heisenbergstr.~1, D-70569 Stuttgart, Germany}

\author{S.~Blanco-Canosa}
\affiliation{Max-Planck-Institut~f\"{u}r~Festk\"{o}rperforschung,
Heisenbergstr.~1, D-70569 Stuttgart, Germany}

\author{M.~Bakr}
\affiliation{Max-Planck-Institut~f\"{u}r~Festk\"{o}rperforschung,
Heisenbergstr.~1, D-70569 Stuttgart, Germany}
\altaffiliation[Present address: ]{Zentrum f\"ur Synchrotronstrahlung, TU Dortmund, D-44221 Dortmund, Germany}

\author{S.~Soltan}
\affiliation{Max-Planck-Institut~f\"{u}r~Festk\"{o}rperforschung,
Heisenbergstr.~1, D-70569 Stuttgart, Germany} \affiliation{Faculty
of Science, Helwan University, 11795-Cairo, Egypt}

\author{M.~Khalid}
\affiliation{Max-Planck-Institut~f\"{u}r~Festk\"{o}rperforschung,
Heisenbergstr.~1, D-70569 Stuttgart, Germany}
\altaffiliation[Present address: ]{Division of Superconductivity and Magnetism, University of Leipzig,
D-04103 Leipzig, Germany}

\author{L.~Mustafa}
\affiliation{Max-Planck-Institut~f\"{u}r~Festk\"{o}rperforschung,
Heisenbergstr.~1, D-70569 Stuttgart, Germany}

\author{K.~Kawashima}
\affiliation{Max-Planck-Institut~f\"{u}r~Festk\"{o}rperforschung,
Heisenbergstr.~1, D-70569 Stuttgart, Germany}

\author{G.~Christiani}
\affiliation{Max-Planck-Institut~f\"{u}r~Festk\"{o}rperforschung,
Heisenbergstr.~1, D-70569 Stuttgart, Germany}

\author{H.-U.~Habermeier}
\affiliation{Max-Planck-Institut~f\"{u}r~Festk\"{o}rperforschung,
Heisenbergstr.~1, D-70569 Stuttgart, Germany}

\author{G.~Khaliullin}
\affiliation{Max-Planck-Institut~f\"{u}r~Festk\"{o}rperforschung,
Heisenbergstr.~1, D-70569 Stuttgart, Germany}

\author{C.~Ulrich}
\affiliation{Max-Planck-Institut~f\"{u}r~Festk\"{o}rperforschung,
Heisenbergstr.~1, D-70569 Stuttgart, Germany}
\affiliation{University of New South Wales, School of Physics, 
Sydney, New South Wales 2052, Australia} \affiliation{The Bragg
Institute, Australian Nuclear Science and Technology Organization,
Locked Bag 2001 Kirrawee DC NSW 2232, Australia}

\author{M.~Le Tacon}
\affiliation{Max-Planck-Institut~f\"{u}r~Festk\"{o}rperforschung,
Heisenbergstr.~1, D-70569 Stuttgart, Germany}

\author{B.~Keimer}
\affiliation{Max-Planck-Institut~f\"{u}r~Festk\"{o}rperforschung,
Heisenbergstr.~1, D-70569 Stuttgart, Germany}

\begin{abstract}
\center\bigskip\thispagestyle{plain}
\begin{minipage}{\textwidth}
\textbf{The electron-phonon interaction is of central importance for the electrical and thermal properties of solids, and its influence on superconductivity, colossal magnetoresistance, and other many-body phenomena in correlated-electron materials is currently the subject of intense research. However, the non-local nature of the interactions between valence electrons and lattice ions, often compounded by a plethora of vibrational modes, present formidable challenges for attempts to experimentally control and theoretically describe the physical properties of complex materials. Here we report a Raman scattering study of the lattice dynamics in superlattices of the high-temperature superconductor $\bf YBa_2 Cu_3 O_7$ and the colossal-magnetoresistance compound $\bf La_{2/3}Ca_{1/3}MnO_{3}$ that suggests a new approach to this problem. We find that a rotational mode of the MnO$_6$ octahedra in $\bf La_{2/3}Ca_{1/3}MnO_{3}$ experiences pronounced superconductivity-induced lineshape anomalies, which scale linearly with the thickness of the $\bf YBa_2 Cu_3 O_7$ layers over a remarkably long range of several tens of nanometers. The transfer of the electron-phonon coupling between superlattice layers can be understood as a consequence of long-range Coulomb forces in conjunction with an orbital reconstruction at the interface. The superlattice geometry thus provides new opportunities for controlled modification of the electron-phonon interaction in complex materials.}
\end{minipage}
\end{abstract}
\maketitle\thispagestyle{empty}\clearpage

The targeted manipulation of the electronic properties of metal-oxide heterostructures and superlattices with atomically precise interfaces is currently at the frontier of materials research \cite{Mannhart_Science2010,Hwang_NatMat2012}. Control parameters including the layer thickness and composition as well as epitaxial strain and gate fields have allowed systematic tuning of many-body phenomena such as ferroelectricity, magnetic order, and superconductivity. The impact of static interfacial lattice distortions on some of these phenomena has already been recognized \cite{Pentcheva_PRL2009,Pauli_PRL2011,Butko_AdvMat2009,Hoppler_NatMat2009}. However, the influence of the dynamical electron-phonon interaction on the electronic properties of artificially layered structures has thus far not been addressed, despite evidence for its crucial role for the phase behavior of bulk transition metal oxides \cite{Dagotto2001,Edwards_AdvPhys2002,Gunnarsson_JPCM2008}.

Here we report a study of the electron-phonon coupling in superlattices composed of the high-temperature superconductor $\rm YBa_2 Cu_3 O_7$ (YBCO) and the metallic ferromagnet $\rm La_{1-x}Ca_{x}MnO_{3}$ (LCMO) with $x=1/3$, which brings together two previously disconnected areas of research. On the one hand, YBCO-LCMO superlattices have served as model systems for the interplay between the antagonistic order parameters of the constituent materials~\cite{Hoppler_NatMat2009,Sefrioui_PRB2003,Holden_PRB2004}, and for interfacial spin~\cite{Stahn2005,Hoffmann_PRB2005,Chakhalian_NatPhys2006} and orbital~\cite{Chakhalian_Science2007} reconstructions. On the other hand, prior research has shown that the electron-phonon interaction and its interplay with electronic correlations that determines the competition between superconducting and charge density wave states in bulk YBCO \cite{Gunnarsson_JPCM2008,Wu_Nature2011,Raichle_PRL2011}, and between correlated metallic and polaronic insulating states in bulk LCMO \cite{Salamon_RMP2001,Edwards_AdvPhys2002,Rivadulla_PRL2006}. In bulk compounds, pressure and chemical substitution offer only limited options to tune this interplay. The outcome of our study identifies the superlattice geometry as a powerful new tool to systematically modify the electron-phonon interaction in complex materials.

\begin{figure}
\includegraphics[width=0.85\linewidth]{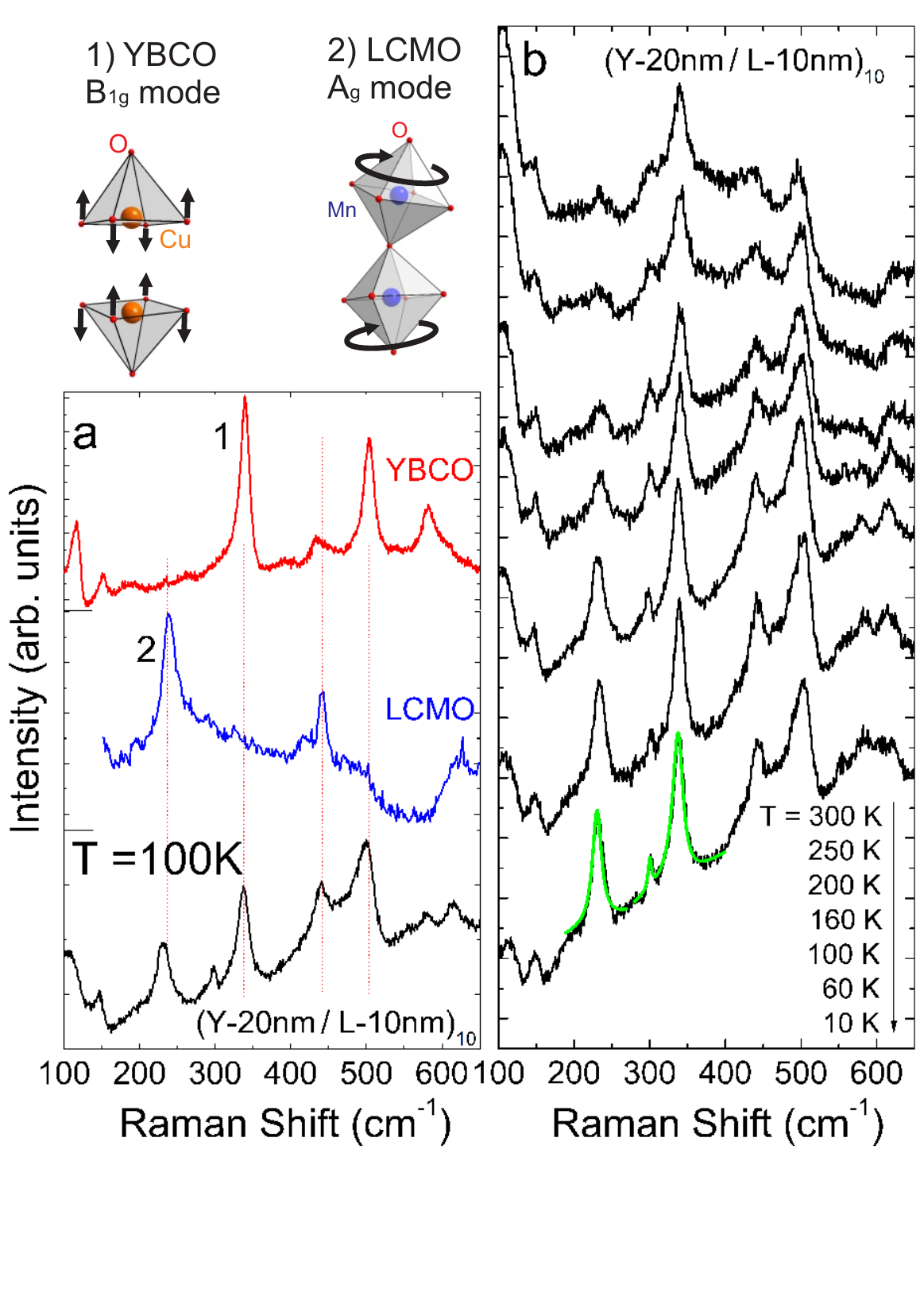}
\caption{(a) Raman spectra of pure YBCO, LCMO, and a (Y-20 nm/L-10 nm)$_{10}$
superlattice at T= 100 K. 1 and 2 denote the peaks corresponding to the B$_{1g}$ 340 cm$^{-1}$ out-of-phase $c$-axis polarized vibration of the planar oxygen atoms in YBCO, and the 230 cm$^{-1}$ A$_g$(2) in-phase rotation of the MnO$_6$ octahedra of LCMO, respectively. The sketches provide pictorial representations of the vibration patterns. b) Temperature dependence of the Raman spectra of a (Y-20 nm/L-10nm)$_{10}$ superlattice in $xx$ polarization. For clarity each spectrum is vertically shifted by a constant offset.}
\label{Rawdata}
\end{figure}

The Raman scattering experiments were performed on superlattices with 10 nm thick LCMO layers, and YBCO layers ranging in thickness from 10 to 50 nm, grown epitaxially on SrTiO$_3$ by pulsed laser deposition following prior work \cite{Chakhalian_NatPhys2006,Chakhalian_Science2007} (see Methods). Figure 1a shows a typical Raman spectrum measured on a superlattice comprising 10 repetitions of 20 nm thick YBCO and 10 nm think LCMO layers, hereafter referred to as (Y-20 nm/L-10 nm)$_{10}$, along with reference spectra of 300 nm thick YBCO and LCMO films.
The reference spectra exhibit two prominent low-energy modes arising from vibrations that modulate the Cu-O-Cu and Mn-O-Mn bond angles in YBCO and LCMO, respectively. Both vibrations have been subjects of extensive prior investigations on bulk samples. The YBCO mode at 340~cm$^{-1}$ (marked ``1'' in Fig. 1a) has B$_{1g}$ symmetry
and is due to a buckling vibration of the CuO$_2$ layers. It shows pronounced superconductivity-induced lineshape anomalies that have provided information about the magnitude and anisotropy of the superconducting energy gap \cite{Friedl_PRL1990,Altendorf_PRB1993,Bakr_PRB2009}. The mode at 230 cm$^{-1}$ in LCMO (marked ``2'' in Fig.~\ref{Rawdata}-a), has A$_g$ symmetry and originates from in-phase rotations of the MnO$_6$ octahedra. Following the literature, we will refer to it as the LCMO A$_g$(2) mode. It shows anomalous behavior at the magnetic ordering transition and has played an important role in research on the colossal-magnetoresistance phenomenon \cite{Irwin_PRB1999,Rivadulla_PRL2006}. In the superlattice spectra, both modes are well separated and can be analyzed individually (Fig. 1a,b). A phonon mode at $\sim 500$ cm$^{-1}$ arising from vibrations of the apical oxygen atoms in YBCO, whose frequency is a sensitive indicator of the oxygen stoichiometry of this material~\cite{Thomsen_SSC1988, Altendorf_PRB1993}, can also be clearly resolved in the superlattice spectra. The Raman-active stretching vibrations of the Cu-O and Mn-O bonds around 450 cm$^{-1}$ overlap strongly in the superlattice spectra, and therefore cannot be quantitatively analyzed.

The LCMO A$_g$(2) and the YBCO B$_{1g}$ phonon profiles in Fig. 1 are much broader than the instrumental resolution and exhibit asymmetric lineshapes, indicating substantial electron-phonon interactions. Figure 2 summarizes the temperature dependence of the phonon frequencies resulting from least-squares fits to standard Fano lineshapes. The behavior of both modes in the superlattices is clearly different from the corresponding reference data on the bulk constituents (also shown in Fig. 2) and exhibits a systematic evolution with YBCO layer thickness.

We first discuss the data on the B$_{1g}$ phonon in YBCO, which are highlighted in Fig. 3 along with prior data on bulk YBCO samples with various oxygen stoichiometries \cite{Altendorf_PRB1993, Bakr_PRB2009}. In both data sets, the phonon energy softens below the superconducting $T_c$, as a result of the loss of electron-phonon decay channels for phonon energies below the superconducting energy gap. The amplitude of the softening decreases progressively with decreasing YBCO layer thickness in the YBCO-LCMO superlattices (Figure~\ref{340_9Samples}c). This is analogous to the effect of decreasing oxygen content in bulk YBCO (Fig.~\ref{340_9Samples}d), where it can be understood as a consequence of the approach to the antiferromagnetic Mott-insulating state~\cite{Altendorf_PRB1993}.

The analogous behavior of the 340 cm$^{-1}$ phonon in both systems indicates a loss of mobile charge carriers in the CuO$_2$ planes with decreasing YBCO layer thickness in the superlattices, as previously suggested based on optical spectroscopy data \cite{Holden_PRB2004}. In the superlattices, however, the change in doping level is not associated with a change in oxygen stoichiometry. This is demonstrated by the frequency of the high-energy apical-oxygen vibration, which depends on the oxygen content in the bulk (Fig.~\ref{340_9Samples}b), but is independent of the layer sequence in the superlattices (Fig.~\ref{340_9Samples}a). The change in doping level in the YBCO layers without chemical substitution can be attributed to charge transfer between the superlattice planes, confirming the conclusions of an earlier x-ray absorption study \cite{Chakhalian_Science2007}. In our data, the superconductivity-induced effects become visible only for superlattices with YBCO layers thicker than 20 nm. This indicates a rather long range scale for the redistribution of charge carriers and the degradation of superconductivity, in accord with prior work ~\cite{Hoppler_NatMat2009,Sefrioui_PRB2003,Holden_PRB2004}.

We now turn to LCMO A$_g$(2) mode. Our reference data for a thick LCMO film (Fig. 2a) reproduce the substantial increase of the frequency of this phonon below the Curie temperature observed in prior Raman scattering work on ferromagnetic LCMO \cite{Irwin_PRB1999}. Since the static magnetostriction at $T_{Curie}$ is much too small to account for this behavior, it has been ascribed to a dynamical modulation of the magnetic exchange interactions through the Mn-O-Mn bond by the vibrational bond-bending motion (see also Supplementary Information)\cite{Irwin_PRB1999}. A hardening of the LCMO A$_g$(2) phonon with decreasing temperature by virtue of this ``spin-phonon coupling'' mechanism has been observed throughout the LCMO phase diagram including the antiferromagnetic Mott-insulator at $x=0$ \cite{Granado_PRB1998} and antiferromagnetic charge-ordered compounds with $x \geq 0.5$ \cite{Liarokapis_PRB1999,Antonakos_MSEB2007}.

Our data on the YBCO-LCMO superlattices, on the other hand, demonstrate a substantial {\it softening} of this mode upon cooling through  $T_{Curie}$ (Figs. 2, 4). Prior Raman scattering experiments have demonstrated softening of higher-energy Mn-O-Mn bond-stretching vibrations at magnetic ordering transitions in LCMO \cite{Granado_PRB1999,Laverdiere_PRB2006} and other manganites~\cite{Issing2010}, and infrared spectroscopy experiments have shown that higher-energy Mn-O-Mn bond-bending modes are sensitive to orbital-ordering phenomena in bulk LCMO \cite{Takazawa_JPSJ2001}. However, the sign reversal of the spin-phonon coupling parameter we have observed in the superlattices is unprecedented to our knowledge. We have directly confirmed that the LCMO A$_g$(2) mode in a thick film of the overdoped charge-ordered antiferromagnet La$_{1/3}$Ca$_{2/3}$MnO$_{3}$ hardens upon cooling below the N\'eel temperature $T_N$ = 150 K (see Supplementary Information). This rules out the possibility that the sign reversal of the spin-phonon coupling is a simple consequence of charge transfer between the YBCO and LCMO layers. Our data thus indicate that the magnetic and orbital order of the LCMO layers in the superlattices is quite different from any of the states that are realized in the bulk. While this is qualitatively consistent with the observation of an ``orbital reconstruction'' at the YBCO-LCMO interface \cite{Chakhalian_Science2007}, the sharp phonon anomaly demonstrates that this reconstruction is not confined to the interfaces, but pervades the entire 10 nm thick LCMO layers.

The low-temperature behavior of the LCMO A$_g$(2) mode is even more surprising. Although the mode energy (and hence the vibration pattern) are very close to the ones observed in bulk LCMO, the mode exhibits a strong anomaly at the {\it superconducting} transition temperature $T_c$ of YBCO (Fig. 2). The amplitude of the superconductivity-induced softening ($\sim 2$\% of the mode energy in the (Y-50 nm/L-10 nm)$_5$ sample) is comparable to the one of the YBCO B$_{1g}$ oxygen vibration, which exhibits by far the strongest superconductivity-induced anomaly of any of the Raman-active phonons in bulk YBCO \cite{Friedl_PRL1990,Altendorf_PRB1993,Bakr_PRB2009}. Figure 4 shows that the frequency shift of the LCMO A$_g$(2) mode goes along with an equally pronounced narrowing of the linewidth, again analogous to the one observed for the  YBCO B$_{1g}$ mode. This implies that the electron-phonon coupling in the YBCO layers is transferred to a vibrational mode in the LCMO layers of the superlattice.

The superconductivity-induced redshift of the LCMO A$_g$(2) mode scales linearly with the thickness of the YBCO layers over a remarkably long range of at least 50 nm (Fig. 4g). This observation indicates that long-range Coulomb interactions, whose influence on the static lattice structure of heterostructures and superlattices composed of covalently bonded transition metal oxides has already been recognized \cite{Butko_AdvMat2009}, play a key role in the lattice dynamics and electron-phonon interaction in such structures as well. A possible explanation of the anomalous behavior of the 230 cm$^{-1}$ mode is thus a small copper-oxygen admixture to the eigenvector induced by the strong Cu-O-Mn bond across the YBCO-LCMO interface \cite{Chakhalian_Science2007}. Long-range Coulomb interactions ensure efficient propagation of this vibrational mode across the thick YBCO layers. Despite its small amplitude, the copper-oxygen component appears to dominate the self-energy of the combined mode, presumably because the vibration pattern of the A$_g$(2) mode in LCMO (Fig. 1) induces bending motions of the Cu-O-Cu bonds in the CuO$_2$ planes. Such vibrations tend to be subject to strong electron-phonon interactions in the cuprates \cite{Gunnarsson_JPCM2008,Raichle_PRL2011}. It is likely that the long-range modification of the electron-phonon interaction contributes prominently to the variation of the electronic properties of YBCO-LCMO superlattices, which involves a similarly large length scale, as noted above.

While detailed calculations are required to further elucidate the origin of the observed long-range transfer of electron-phonon coupling, the data at hand already demonstrate that epitaxial superlattices offer novel opportunities to generate vibrational modes that do not exist in the bulk, and to systematically modify their properties through the layer geometry. This provides a powerful new tool to explore and control the electron-phonon interaction in transition metal oxides at ambient pressure and without introducing chemical disorder.

\bigskip
\noindent\textbf{Methods}
High quality YBCO-LCMO superlattices were grown by pulsed laser deposition (PLD) on SrTiO$_{3}$ substrates with (001) orientation~\cite{Soltan_PRB2004}. In this configuration, the $c-$axis of YBCO is out of plane (that is, the CuO$_2$ planes are parallel to the substrate surface), as determined by X-ray pole figures and polarization dependent measurements of the Raman-allowed phonon modes. The superconducting and ferromagnetic transition temperatures were measured using a Vibrating Sample Magnetometer (VSM-SQUID). The corresponding transition temperatures listed in Table I are consistent with previous work~\cite{Sefrioui_PRB2003,Pena_PRB2004}. More details are provided in the Supplementary Information.

\begin{table}[b]
\begin{ruledtabular}
\begin{tabular}{c|cccc}
   Sample       &      YBCO          &     LCMO           &     $T_{c}$       &          $T_{Curie}$ \\
                &      Thickness      &   Thickness       &                    &                      \\
\hline
bulk LCMO           &       0 nm           &      300 nm  &       -             & 275 K                \\

(Y-10 nm/L-10 nm)$_{15}$ & 10 nm & 10 nm & 45 K & 230 K\\

(Y-20 nm/L-10 nm)$_{10}$ & 20 nm & 10 nm & 60 K & 220 K\\

(Y-30 nm/L-10 nm)$_{7}$ & 30 nm & 10 nm & 82 K & 185 K\\

(Y-50 nm/L-10 nm)$_{5}$ & 50 nm & 10 nm & 82 K & 230 K\\

 bulk YBCO           &       300 nm            &      0 nm &       90 K             &  -                \\
\end{tabular}
\end{ruledtabular}
\caption{\label{tab:samples} Summary of the properties of the samples used in this work. The columns denote the
YBCO and LCMO thicknesses and the superconducting and ferromagnetic transition
temperatures, respectively. The total thickness of the samples was kept between 250-300 nm in order to avoid contributions of the substrate to the Raman signal.}
\end{table}

The Raman scattering experiments were performed in backscattering geometry using a Horiba-Jobin-Yvon T64000 micro-Raman spectrometer equipped with a nitrogen cooled CCD camera as detector. The 514.5 nm and 488.0 nm Ar$^{+}$ laser lines were used for the measurements, and the reported results were found to be independent of the chosen wavelength. The incident laser beam was focussed to a 10 $\mu m$ spot using a $\times 50$, long working distance objective on the sample which was placed on the cold finger of a He-cooled cryostat. The laser power was kept below 1.5 mW to avoid laser-induced heating. The polarization of the electric field of the incident and scattered light was kept parallel to the crystallographic $a-$axis of the YBCO layer. The strain induced by the cubic substrate prevents us from distinguishing between the crystallographic $a$ and $b$ axes of YBCO unit cell. Hence, we use the tetragonal notation D$_{4h}$ (B$_{1g}$, A$_{1g}$) instead of the orthorhombic D$_{2h}$ (A$_g$) to assign the Raman-allowed modes of YBCO.

The Raman data were fitted using the Fano lineshape $I(\omega) = A [(q+\varepsilon)^{2} / (1+\varepsilon^{2}) ]$, where $\varepsilon= (\omega-\omega_{0})/\Gamma$, $\omega_{0}$ and $\Gamma$ are the phonon frequency and linewidth (half width at half maximum), respectively. \textit{A} is a proportionality constant and \textit{q} is the Fano asymmetry parameter~\cite{Altendorf_PRB1993,Friedl_PRL1990,Bakr_PRB2009}.

\noindent\textbf{Acknowledgements}
Part of this research project has been supported by the European Commission under the 7th Framework Programme Marie Curie action SOPRANO project (Grant No. PITNGA-2008-214040).

\noindent\textbf{Author contributions}
N. D. and S. B.-C. contributed equally to this work.
L.~M., K.~K., G.~C., S.~S. and H.-U.~H. provided the superlattices.
N.~D. and S.~B.-C. performed the sample characterization and the Raman experiments.
N.~D., S.~B.-C., M.~L.~T. analyzed the data.
M.~B., M.~K., and C.~U. participated in the Raman measurements and analysis.
G.~K. contributed to the discussion and interpretation of the
results.  M.~L.~T. and B.~K. wrote the manuscript.
M.L.T., C.~U.~and B.~K. supervised the project.

\noindent\textbf{Author Information} Correspondence and requests for materials should be addressed to
M.~L.~T. (\href{mailto:m.letacon@fkf.mpg.de}{m.letacon@fkf.mpg.de}) or B.~K.\ (\href{mailto:v.b.keimer@fkf.mpg.de}{b.keimer@fkf.mpg.de}).


\newpage

\begin{figure*}
\includegraphics[width=0.9\linewidth]{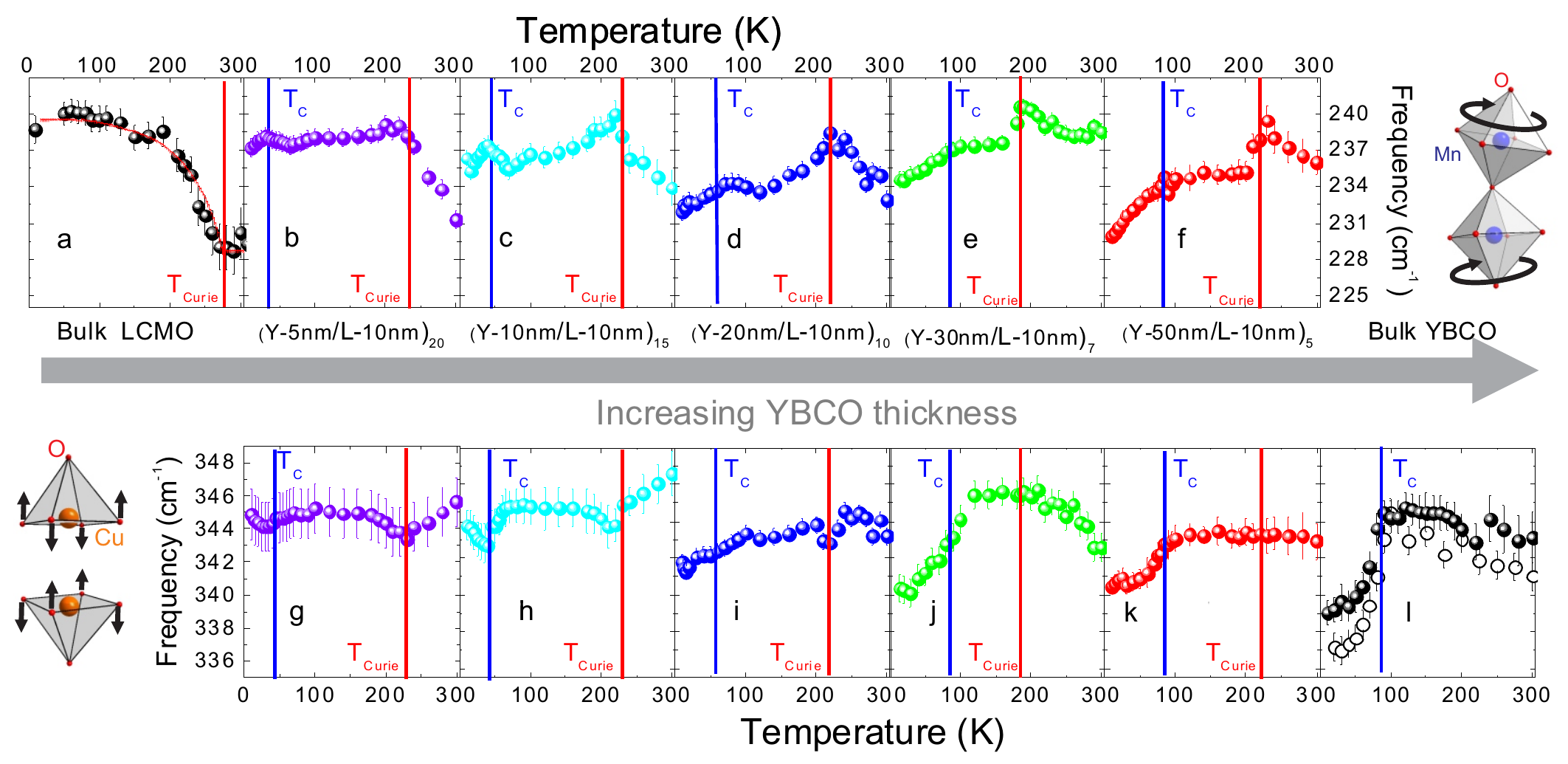}
\caption{(a-e) Temperature dependence of the A$_g$(2) LCMO phonon frequency and (f-j) of the B$_{1g}$ YBCO phonon
frequency of the different samples studied here. Panel j shows the temperature dependence of the B$_{1g}$ mode frequency measured on a 300 nm ~thick film (black spheres) and single crystal (open circles) from Ref. \onlinecite{Bakr_PRB2009}. The red line in panel a is the result of a fit to the model described in Ref. \onlinecite{Irwin_PRB1999} (see Supplementary information). In every panel, the error bars reflect the accuracy of the fitting procedure.}
\label{230cm_samples}
\end{figure*}

\begin{figure}
\includegraphics[width=0.95\linewidth]{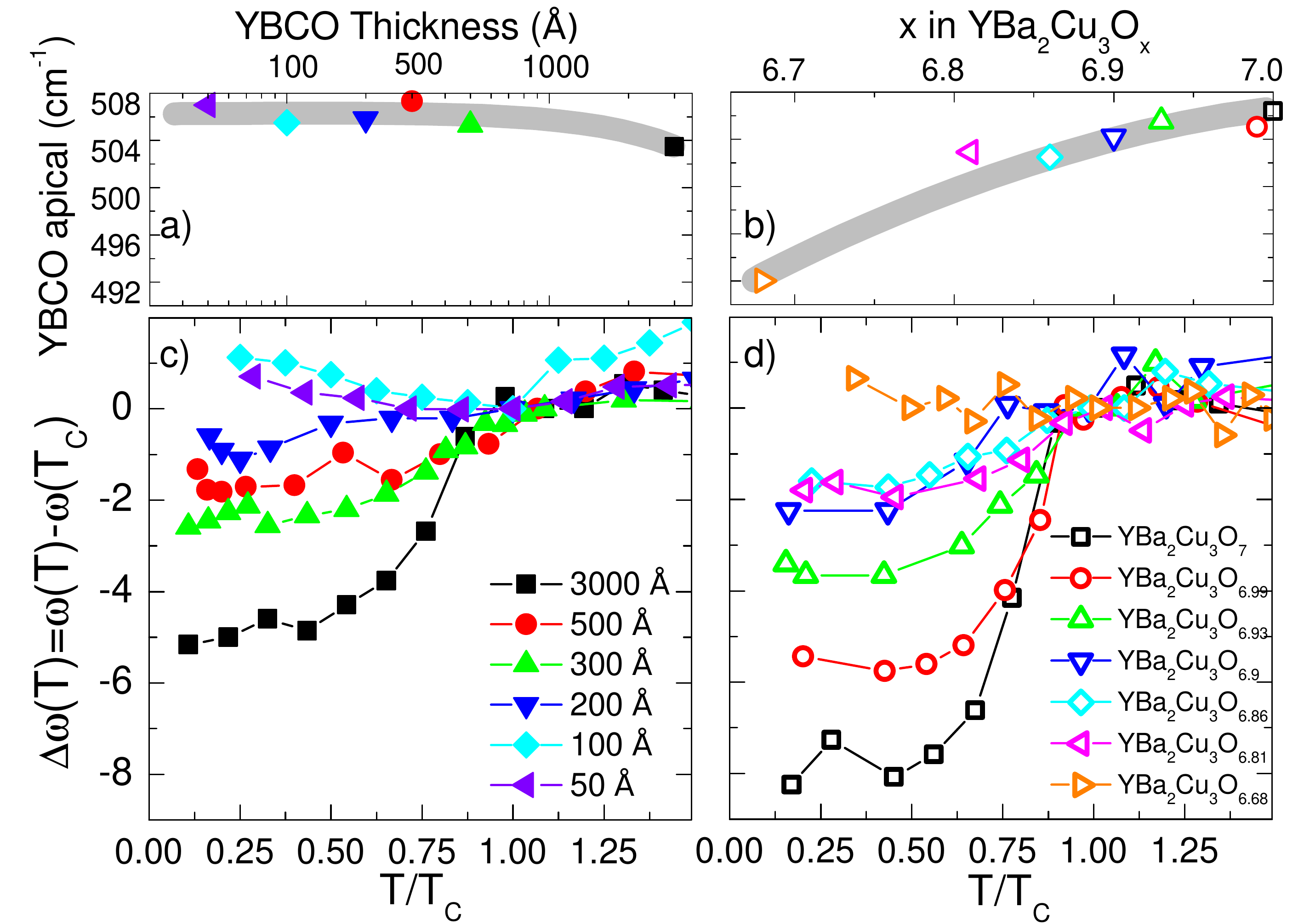}
\caption{(a) Frequency of the A$_{1g}$ apical O mode of YBCO as a function of the layer thicknesses.  b) O-doping dependence of the frequency of the A$_{1g}$ apical O mode in YBa$_2$Cu$_3$O$_{6+x}$ single crystals (from Ref. \onlinecite{Altendorf_PRB1993}). c) Superconductivity-induced renormalization of the 340 cm$^{-1}$ B$_{1g}$ phonon for the various YBCO-LCMO superlattices studied here. d) Doping dependence of the superconductivity-induced renormalization of the 340 cm$^{-1}$ B$_{1g}$ phonon in YBa$_2$Cu$_3$O$_{6+x}$ single crystals (from Ref. \onlinecite{Altendorf_PRB1993}).}
\label{340_9Samples}
\end{figure}

\begin{figure}
\includegraphics[width=0.75\linewidth]{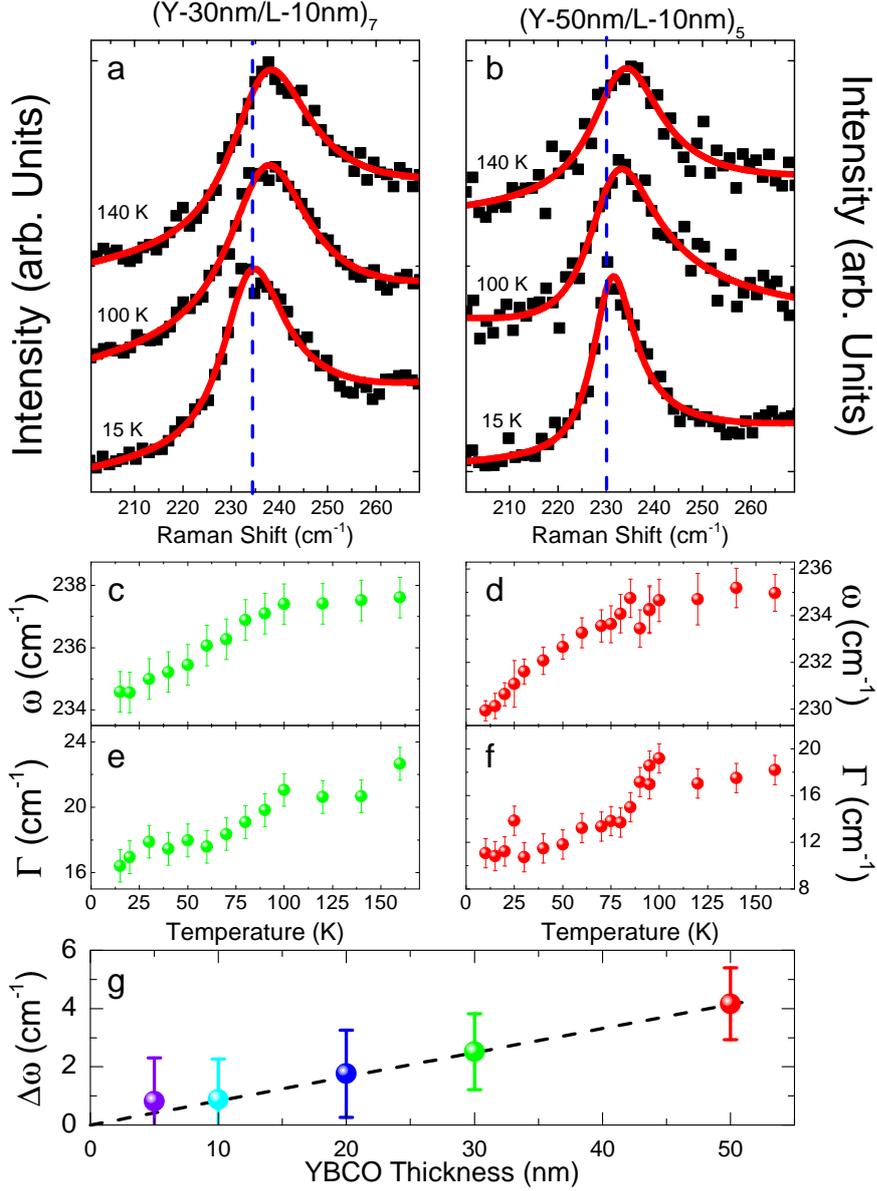}
\caption{a) LCMO A$_g$(2) phonon at T = 140, 100 and 15 K in a (Y-30 nm/L-10 nm)$_{7}$ superlattice. The black squares are the experimental data points, and the red lines are the results of fits to Fano profiles (see Methods). b)  LCMO A$_g$(2) phonon at 140, 100 and 15 K in a (Y-50 nm/L-10 nm)$_{7}$ superlattice. c) Temperature dependence of the frequency of the LCMO A$_g$(2) mode in a (Y-30 nm/L-10 nm)$_{7}$ superlattice.  d)  Temperature dependence of the frequency of the LCMO A$_g$(2) mode in a (Y-50 nm/L-10 nm)$_{7}$ superlattice. e) Temperature dependence of the linewidth of the LCMO A$_g$(2) mode in a (Y-30 nm/L-10 nm)$_{7}$ superlattice. f) Temperature dependence of the linewidth of the LCMO A$_g$(2) mode in a (Y-30 nm/L-10 nm)$_{7}$ superlattice. g) YBCO thickness dependence of the A$_g$(2) LCMO phonon softening through $T_c$ in our superlattices. In panels c through g, the error bars reflect the accuracy of the fitting procedure.}
\label{LCMO_TC}
\end{figure}

\end{document}